\documentclass[twocolumn,numbers,amsmath,amssymb]{revtex4}
\usepackage{graphicx}
\usepackage{dcolumn}
\usepackage{bm}
\begin{document}

\title{Magnetoelectric Coupling and Electric Control of Magnetization in Ferromagnet-Ferroelectric-Metal Superlattices}

\par
\author{Tianyi Cai,$^{1,3}$ Sheng Ju,$^{2}$  Jaekwang Lee,$^{1}$ Na Sai,$^{1}$ Alexander A. Demkov,$^{1}$ Qian Niu,$^{1}$ Zhenya Li,$^{2}$ Junren Shi,$^{3}$ and Enge Wang$^{3}$  } \affiliation{$^{1}$ Department of Physics, The University of
Texas, Austin 78712 \\ $^{2}$ Department of Physics and Jiangsu
Key Laboratory of Thin Films, Soochow University, Suzhou, 215006, China \\
$^{3}$ Institute of Physics and ICQS, Chinese Academy of Sciences,
Beijing 100080, China}

\begin{abstract}
Ferromagnet-ferroelectric-metal superlattices are proposed to
realize the large room-temperature magnetoelectric effect. Spin
dependent electron screening is the fundamental mechanism at the
microscopic level.  We also predict an electric control of
magnetization in this structure. The naturally broken inversion
symmetry in our tri-component structure introduces a
magnetoelectric coupling energy of $P M^2$. Such a magnetoelectric
coupling effect is general in ferromagnet-ferroelectric
heterostructures, independent of particular chemical or physical
bonding, and will play an important role in the field of
multiferroics.

\end{abstract}

 \maketitle

Ferroelectricity and ferromagnetism are important in many
technological applications and the quest for multiferroic
materials, where these two phenomena are intimately coupled, is of
significant technological and fundamental interest
\cite{fie05,spaldin,eer06,che07,ram07,tok07}. In general,
ferroelectricity and ferromagnetism tend to be mutually exclusive
or interact weakly with each other when they coexist in a
single-phase material\cite{spaldin}. Increasing the spin-orbit
interaction of the electrons~\cite{ref1} or strategically
designing for magnetic and electric phase
control~\cite{tok07,fen06} may enhance the magnetoelectric (ME)
coupling effect in a single phase multiferroic material. Practical
applications of the ME effect, however, remain hindered by the
small electric polarization and low Curie
temperature\cite{eer06,che07,ram07}.

Artificial composites of ferroic materials may enable the
room-temperature ME effect, since both large and robust electric
and magnetic polarizations can persist to room temperature. Two
types of ME coupling at a ferromagnet(FM)-dielectric interface
have been reported, one employing the mechanical
interaction\cite{zhe04,zav05} or chemical bonding~\cite{duan06}
and the other mediation by carriers (screening charges)
~\cite{zhang99, nicola07, duan08, Maruyama}. {The role of electrostatic screening in ferroelectric capacitors has been studied by macroscopic models~\cite{batra}. Recently, {\it ab-initio}
studies of nanoscale ferroelectric(FE)
capacitors~\cite{jun03,sai05} and FE tunnel
junctions~\cite{tsy06,zhu05,ju07} have further confirmed that
electrostatic screening is the fundamental mechanism at the
FE/normal metal (NM) interface. In this letter we propose a
strategy of achieving robust ME coupling in a tri-component
FM/FE/NM superlattice. The additional magnetization, caused by
spin-dependent screening\cite{zhang99,nicola07}, will accumulate
at each FM/FE interface. Due to the broken inversion symmetry
between the FM/FE and NM/FE interfaces, there would be a net
additional magnetization in each FM/FE/NM unit cell, unlike the
symmetric structures discussed in the previous work. The addition
of magnetization in this superlattice will result in a large
global magnetization.

\begin{figure}
\includegraphics{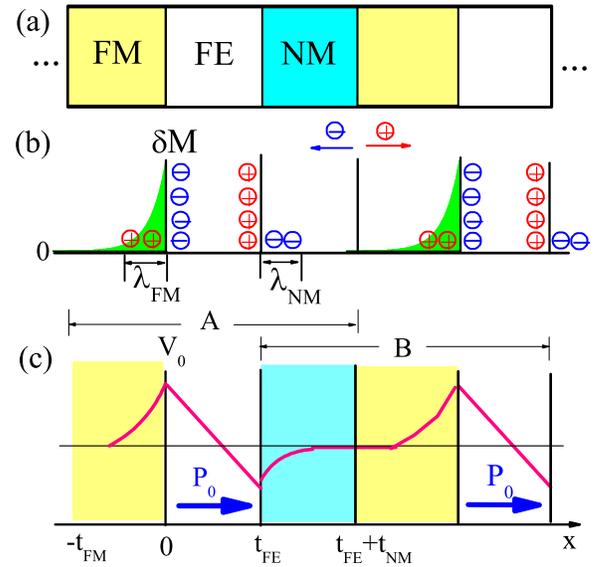}
\caption{\label{fig:epsart} (Color online). (a) Schematic
illustration of FM/FE/NM tri-component superlattice. (b) The
distribution of charges and induced magnetization (green)
calculated by our theoretical model. A and B are two different
choices of the unit cell. The directions of arrows indicate the
motions of positive (red) and negative (blue) charges across the
boundary of the unit cell A. (c) Electrostatic potential profile.
}
\end{figure}
The tri-component superlattice is illustrated in Fig.1 (a). When
the FE layer is polarized, surface charges are created. These
bound charges are compensated by the screening charge in both FM
and NM electrodes. In the FM metal, the screening charges are spin
polarized due to the ferromagnetic exchange interaction. The spin
dependence of screening leads to additional magnetization in the
FM electrode as illustrated in Fig.1(b). If the density of
screening charges is denoted as $\eta$ and the spin polarization
of screening charges as $\zeta$, we can directly express the induced
magnetization per unit area as
\begin{equation}
\Delta M = \frac{\eta }{e} \cdot \zeta \mu _B .
\end{equation}
As this effect depends on the orientation of the electric
polarization in FE, the magnetoelectric (ME) coupling is expected.

Before starting specific calculations, let's consider two simple
cases. (1) In an ideal capacitor where all the surface charges
reside at the metal (FM or NM)/FE interfaces, the density of
screening charge $\eta$ reaches its maximum value, $\eta=P_0$,
where $P_0$ is the spontaneous polarization of the FE. This
results in a large induced magnetization
($\frac{P_0}{e}\cdot\zeta\mu_B$). (2) In half metals, there is
only one type of carriers that can provide the screening. If a
half metal is chosen to be the FM electrode, the screening
electrons will be completely spin-polarized. In this case, a large
induced magnetization is also expected, $\Delta M$=
$\frac{\eta}{e}$$\mu_B$.

\textit{\textbf{Induced Magnetization from screening charges.}}
For simplicity, we will first consider the case of zero bias, as
illustrated in Fig.1c. Here, the following assumptions are made.
(1)The difference in the work function between FM and NM is
ignored. (2)To screen the bound charges in FE, the charges in
metal electrodes will accumulate at the FM/FE side, and there is a
depletion at the NM/FE side. In this process, the total amount of
charge is conserved, however, the spin density is not conserved
because of the ferromagnetic exchange interaction in the FM metal.

As shown in Fig.1, the \textit{local} induced magnetization,
defined as $\delta M(x)=[\delta n^\uparrow (x)-\delta n^\downarrow
(x)]\mu_B$, is a function of distance from the interface $x$.
Here, $\delta n^\sigma(x)$ is the density of the \textit{induced}
screening charges with spin $\sigma$. Zhang~\cite{zhang99}
considered the FM/dielectric interface within the linearized
Thomas-Fermi model and derived two coupled equations relating the
local induced magnetization $\delta M (x)$ and screening potential
$V_0(x)$,
\begin{equation}
\left\{ {\begin{array}{*{20}c}
   {\delta M(x) =  - \frac{{M_0/\mu_B}}{{1 + JN_0 }}eV_0 (x)}  \\
   {\frac{{d^2 V_0 (x)}}{{dx^2 }} = \frac{1}{{\lambda _{FM}^2 }}V_0 (x)}  \\
\end{array}} \right.
\end{equation}
The screening length in the FM electrode is defined as
$\lambda_{FM} = \left\{ {\frac{{e^2 N_0 }}{\varepsilon_0
}\frac{{N_0 + JN_0^2  - J(M_0/\mu_B)^2 }}{{1 + JN_0 }}} \right\}^{
- 1/2}$, where $N_0=N ^ \uparrow + N ^ \downarrow $ is the total
density of states, $M_0=(N^\uparrow-N^\downarrow$)$\mu_B$ can be
thought of as the spontaneous magnetization, $\varepsilon_0$ is
the vacuum dielectric constant and $J$ is the strength of the
ferromagnetic exchange coupling in the FM layer.

We solve the above equations for our unit cell and obtain
\begin{equation}
V_0 (x) = \left\{ {\begin{array}{*{20}c}
   {\frac{{\eta \lambda _{FM}   }}{{\varepsilon_0} }e^{x/\lambda _{FM}},} & {-t_{FM}\le x \le 0;}  \\
   { - \frac{{\eta \lambda _{NM}  }}{{\varepsilon_0 }}e^{ - (x - t_{FE})/\lambda _{NM} },} & {t_{FE} \le x \le t_{FE}+t_{NM}.}  \\
\end{array}\begin{array}{*{20}c}
   {}  \\
   {}  \\
\end{array}} \right.
\end{equation}
where $\eta$ is the density of screening charges, $\lambda
_{FM(NM)}$ is the screening length of FM(NM) electrode, and
$t_{FM}$, $t_{FE}$ and $t_{NM}$ are the thickness of FM, FE and NM
layer, respectively. From above equations, we see that the local
induced magnetization $\delta M(x)$ decays exponentially away from
the FM/FE interface. This distribution is identical to that of
screening charges, because in our model the effective interaction
$J$ in FM is assumed to be a constant. The total induced
magnetization $\Delta M$ can be calculated by integrating $\delta
M(x)$ over the FM layer,
\begin{equation}
\Delta M=\int_{\text{FM layer}} \delta M(x)=-\frac{{\eta M_0 /e
}}{{N_0+ JN_0^2-J(M_0/\mu_B)^2}}.
\end{equation}
The effective spin polarization of screening electrons can then be
written as
\begin{equation}
 \zeta=- \frac{{M_0 }}{{N_0 + JN_0^2 - J(M_0/\mu_B)^2 }}.
\end{equation}

We have considered the induced magnetization in FM/FE/NM
tri-component superlattice with several FM electrodes, i.e., Fe,
Co, Ni and CrO$_2$. Detailed parameters and calculated values of
$\Delta M$ are listed in Table I. The magnitude of $\Delta M$ is
found to depend strongly on the choice of the FM and FE. Among the
normal FM metals (Ni, Co and Fe), the largest $\Delta M$ is
observed in Ni for its smallest $J$ and highest spontaneous spin polarization $M_0/\mu_BN_0$. On
the other hand, we also predict a large $\Delta M$ for the 100\% spontaneous spin polarization in half-metallic CrO$_2$.

\begin{table*}
\caption{\label{tab:table1}Parameters extracted from band
structures of Ni, Co, Fe from Ref.\cite{zhang99}, CrO$_2$ from
Ref.\cite{bra97}. Here, $\Delta M$ is the value at $V_a$=$V_C$,
where $V_a$ is the applied bias and $V_C$ is the coercive bias. }
\begin{ruledtabular}
\begin{tabular}{cccccccc}
FM  & $J$ (eV$\cdot$nm$^{3}$) & $N_0$ (eV$^{-1}$$\cdot$nm$^{-3}$)  & $M_0/\mu_BN_0$  & $\lambda_{FM}$ (\AA) & $\Delta M$ ($\mu_B$$\cdot$nm$^{-2}$)& $\tau$ (G cm/V) \\
\hline
Ni & 0.65 & 1.74 & -79.3\% & 0.9 & -0.280 & 0.015 \\
Co & 1.25 & 0.89 & -58.4\% & 1.5 & -0.126 & 0.004 \\
Fe & 2.40 & 1.11 & 56.8\% & 1.3 & 0.078& 0.003 \\
CrO$_2$ & 1.8 & 0.69 & 100\% & 1.7 & 0.323& 0.010 \\
\end{tabular}
\end{ruledtabular}
\end{table*}

To confirm the validity of our model, we perform first-principles
calculations of the Fe/FE/Pt superlattice~\cite{superlattice}. The
calculations are within the local-density approximation to
density-functional theory and are carried out with
VASP~\cite{DFT}. We choose BaTiO$_3$ (BTO) and PbTiO$_3$ (PTO) for
the FE layer. Starting from the ferroelectric P4mm phase of BTO
and PTO with polarization pointing along the superlattice stacking
direction, we perform structural optimization of the multilayer
structures by minimizing their total energies. The in-plane
lattice constants are fixed to those of the tetragonal phase of
bulk FEs. Fig. 2 shows the calculated induced magnetic moment
relative to that of bulk Fe near the Fe/BTO interface when the
polarization in BaTiO$_3$ points toward the Fe/BTO interface. It
is evident that the induced moments decay exponentially as the
distance from the interface increases.  This result is in line
with our model for the magnetization accumulation in the FM at the
FM/FE interface. A numerical fitting of the exponential function
yields a screening length of $\sim0.7$ \AA~ for the
Fe/BaTiO$_3$/Pt structure. This value is comparable to the
screening length parameters calculated using the theoretical model
as shown in Table I.
\begin{figure}
\includegraphics[width = 9 cm]{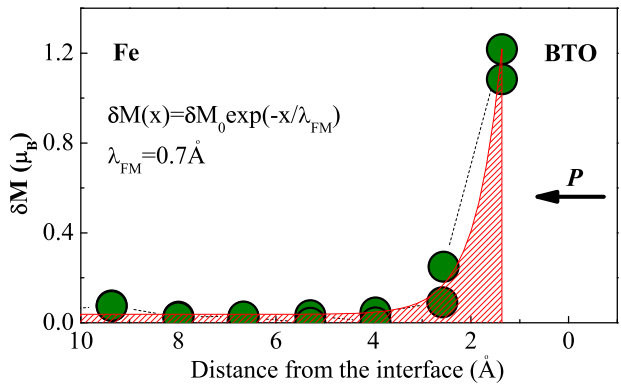}
\caption{\label{fig:epsart} (Color online). Layer-resolved induced
magnetic moment of Fe near the interface between Fe and BaTiO$_3$
in the Fe/BaTiO$_3$/Pt superlattice. Solid line is the fitted
exponential function for the induced moment as a function of the
distance from the interface.}
\end{figure}

Using our theoretical model we also calculate the ME coupling
coefficient $\tau$, which is defined as the ratio of the
magnetization change 2$\mu_0 \Delta M/\lambda_{FM}$ to the
coercive field $V_C/(l*t_{FE})$, where $V_C$ is the coercive
voltage and $l$ is the number of the unit cell. The $\tau$ values
listed in Table I reach as large as 0.015 G cm/V in Ni. For
comparison, $\tau$ of about 0.01 G cm/V arising from the chemical
bonding between Fe and Ti atoms is predicted for the Fe/BaTiO$_3$
bilayer \cite{duan06}. The ME coefficient measured in epitaxial
BiFeO$_3$-CoFe$_2$O$_4$ columnar nanostructures \cite{zav05} is
also of 0.01 G cm/V. We should point out that in our calculation
the coercive field is assumed to be 200 kV/cm, if we choose the
coercive field of 10 kV/cm same as Ref. \cite{duan06}, $\tau$ will
be 20 times larger than those listed in Table I. Therefore, the ME
effect arising from spin-dependent electron screening in FM/FE/NM
tri-component superlattice can be \textit{much larger} than in
other composite multiferroics.

What is the source of this large ME effect? In fact, the
magnetoelectric effect discussed in this letter is not the usual
bulk magneto-electric coupling at all. Spontaneous electric
polarization in FE results in the induced surface charge. In turn,
this produces the screening charges of density $\eta$. These
screening charges are polarized with the polarization $\zeta$.
Therefore, it is the amount of screening charges and polarization
that determine the magnitude of the ME effect ($\Delta M$ and
$\tau$). If we expand the induced magnetization $\Delta M$ in
Eq.(4) as power series in order parameters $P_0$ and $M_0$
(spontaneous polarization and magnetization), we obtain
\begin{equation}
\Delta M \propto P_0M_0 + \text{high-order terms in P$_0$ and
M$_0$}
\end{equation}
The higher order terms in Eq.(3) vanish exactly in the following
limiting case: the screening length $\lambda_{FM}\rightarrow 0$
and spin polarization $\zeta\rightarrow\pm 100\%$. In general, the
leading term in Eq.(6) is linear, which is consistent with the
computational result of Ref.\cite{nicola07}. It is also clear that
this effect depends on the magnetization of the ferromagnetic
metal.

First-principle calculations confirm the central conclusion that
the ME coupling in the tri-component system is linear in
polarization of FE. We compare a superlattice with BTO and that
with PTO. The induced magnetization difference is 3.6 times larger
in the the case of PTO. This ratio is almost exactly that of the
bulk spontaneous polarization of BTO and PTO.

\textit{\textbf{Electric control of magnetization.}} So far, we
have discussed the magnetoelectric coupling effect in the case of
no external bias. A natural question is what happens to $\Delta M$
when external bias $V_a$ is applied. In this case, the electric
polarization $P$ will have two parts: the spontaneous polarization
$P_0$ and induced polarization. The equation determining $P$ is
obtained by minimizing the Free energy. From the continuity of the
normal component of the electric displacement, we find equation
relating $\eta$ and $P$: $\eta=(\frac{P\cdot t_{FE}
}{\varepsilon_{FE}}+\frac{V_a}{l})/(\frac{\lambda_{FM}+\lambda_{NM}}{\varepsilon_0}+\frac{t_{FE}}{\varepsilon_B})$.
Here, $\varepsilon_{FE}$ is the dielectric constant of the FE
layer. These two equations need to be solved self-consistently.
The value of $\eta$ at a given bias can then be calculated and the
induced magnetization $\Delta M$ is given by Eq.(4).

The free energy density $ F$ includes contributions from the FE
layer, FM layer and FM/FE interface and takes the form
\begin{equation}
F = \frac{{t_{FE}  \cdot F(P ) + t_{FM}  \cdot F(M ) + F_I(P ,M )
}}{{t_{FE}  + t_{FM}  + t_{NM} }}.
\end{equation}
$M$ is the magnetization of the bulk ferromagnet, and here
$M$=$M_0$ because of zero external magnetic field. The interface
energy $F_I (P, M)$ is the sum of the electrostatic energy and
magnetic exchange energy of the screening charges
\begin{equation}
F_I (P ,M ) = \frac{{(\lambda _{FM}  + \lambda _{NM}
)}}{{2\varepsilon_0 }}\eta ^2  + \frac{{J }}{{2\mu _B ^2 }}(M  +
\Delta M) \cdot \Delta M.
\end{equation}
For FE, the free energy density $F(P )$ can be expressed as $ F(P
) = F_{P}  + \alpha_{P} P^2 + \beta_{P} P ^4  + \int_0^{P } {E_B
dP}$, where $F_P$ is the free energy density in the unpolarized
state. $\alpha_P$ and $\beta_P$ are the usual Landau parameters of
bulk ferroelectric. $E_B$ is the depolarization field in the FE
film. Similarly, $F(M)$ can be expanded as a series in order
parameter $M$, i.e., $F(M ) = F_{M}  + \mu_{M} M ^2  + \nu_{M} M
^4$, where $F_M$ is the free energy density of bulk ferromagnet,
$\mu_{M}$ and $\nu_{M}$ are the Landau parameters of bulk
ferromagnet. The calculated induced magnetization as function of
applied bias is shown in Fig.3. Clearly, the electrically
controllable magnetization reversal is realized.
\begin{figure}
\includegraphics[width = 9 cm]{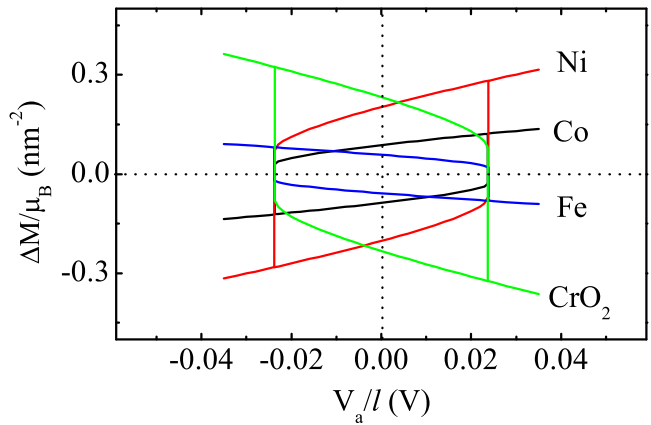}
\caption{\label{fig:epsart} (Color online)  $\Delta M$ versus
$V_a/l$ for different ferromagnetic metal electrodes. $V_a$ is the
applied bias and $l$ is the number of the unit cell. Here, the
thickness of FE layer is 3 nm. However, a thicker FE layer
can be used to avoid the possible electron tunneling effect.} 
\end{figure}

To discuss the macroscopic properties of the electric control of
magnetization, we analyze the magnetoelectric coupling energy in
our tri-component superlattice. For the macroscopic average
polarization to be represented by the electric polarization
obtained for a unit cell, this cell needs to be chosen with
special care \cite{niu}. Therefore, in the following calculation
of total free energy, unit cell $\bf B$ in Fig.1(b) is chosen, and
\begin{equation}
 \bar P = \frac{{P\cdot t_{FE}  + \eta (t_{FM}  + t_{NM} )}}{{t_{FE}  + t_{FM}  + t_{NM} }}.
\end{equation}
The macroscopic average magnetization $\bar M$ is
\begin{equation}
 \bar M = \frac{{M\cdot t_{FM}  + \Delta M}}{{t_{FE}  + t_{FM}  + t_{NM} }}.
\end{equation}
Considering the lowest order term of the magnetoelectric coupling,
$\bar P$ and $\bar M$ can be expanded as $\bar P= c_p P+c_p'PM^2;
\bar M= c_m M+ c_m'PM$. Therefore, the total free energy (Eq.(7))
can be expressed as the power series of $\bar P$ and $\bar M$, $
 F(\bar P,\bar M) = F_0 + \alpha \bar P^2 +\beta \bar
P^4 + \mu \bar M^2 +\nu \bar M^4  + \chi \bar P\bar M^2+\cdots.$
We would like to point out that biquadratic ME coupling $\bar
P^2\bar M^2$ is easily achievable, but is usually weak and is not
electrically controllable. However, because of the naturally
broken inversion symmetry, the large ME coupling $ \bar P\bar M^2$
is possible in our tri-component structure.

The ME coupling in FM/FE/NM superlattices may be observed
experimentally and may have practical applications. Though the net
additional magnetization of each FM/FE/NM unit cell is small,
stacking several of them in a superlattice will result in a large
overall magnetization. From Eq.(10), with a thinner metallic
electrode, the ME effect will be larger, as long as the thickness
of metallic electrodes are larger than the screening length, which
is easy to achieve.

To summarize, expanding upon the theory of spin-dependent
screening\cite{zhang99}, we develop a theory of additional
magnetization in tri-component superlattice. We show that the
additional magnetization can be electrically controlled, and is
linear in FE polarization. The latter can be switched if the coercive voltage of the
ferroelectric is reached.  We demonstrate that an asymmetric FM/FE/NM structure has practical
advantages over previously discussed symmetric structure.

{\it Acknowledgments:} TC and QN were supported by DOE (DE-FG03-02ER45985), NSF(DMR0906025), Welch Foundation (F-1255), and NSFC (10740420252). SJ were supported by K. C. Wong Education Foundation, Hong Kong and China
Postdoctoral Science Foundation. JL,NS and AAD were supported by
the Office of Naval Research (N000 14-06-1-0362) and
Texas Advance Computing Center.

\end{document}